\documentclass[11pt,a4paper]{article} 

\usepackage{anysize}
\usepackage[format = hang]{caption}
\usepackage{amsmath,amsthm,verbatim,amssymb,amsfonts,amscd, graphicx}
\usepackage{graphics,natbib}
\usepackage{titlesec,footmisc}
\usepackage{hyperref} 

\usepackage{listings}
\usepackage{booktabs} 

\theoremstyle{plain}

\theoremstyle{definition}

\def\Bet{{\rm Beta}}
\def\be{{\rm be}}
\def\Be{{\rm Be}}

\def\E{{\rm E}}
\def\Var{{\rm Var}}

\def\midd{\,|\,}

\def\hatt{\widehat}
\def\arr{\rightarrow}

\def\half{\hbox{$1\over 2$}}

\def\Dir{{\rm Dir}}

\def\beq{\begin{eqnarray}}
\def\eeq{\end{eqnarray}}

\def\beqn{\begin{eqnarray*}}  
\def\eeqn{\end{eqnarray*}}

\def\E{{\rm E}}
\def\Var{{\rm Var}}

\def\Pr{P}

\def\arr{\rightarrow}
\def\hatt{\widehat}
\def\tilda{\widetilde}

\def\half{\hbox{$1\over2$}}

\def\obs{{\rm obs}}
\def\midd{\,|\,}

\def\cc{{\rm cc}}

\titleformat{\section}{\normalfont\large\sc\centering}{\thesection}{1em}{}
\titleformat{\subsection}[runin]{\normalfont\large\bfseries}{\thesubsection}{1em}{}
\numberwithin{equation}{section} 
\renewenvironment{abstract}
               {\list{}{\rightmargin\leftmargin}%
                \item[\text{\hspace{10mm}\sc Abstract.}]\relax}
               {\endlist}

\begin{document}

\def\heute{August 2026}

\begingroup
\begin{centering} 

\Large{\bf How many cards, until the first ace: \\
    variations, extensions, lachrymae, confidence, Dirichlets}\\[0.8em]
\large{\bf Nils Lid Hjort} \\[0.3em] 
\small {\sc Department of Mathematics, University of Oslo} \\[0.3em]
\small {\sc {\heute}}\par
\end{centering}
\endgroup


\begin{abstract}
\small{From a deck of cards, how many cards do I need
  to draw, until the first ace? I identify the distribution
  for this waiting time $T$, and its satisfyingly nice
  expected value $\E\,T=(N+1)/(n+1)$, with $N$ the number
  of cards and $n$ the number of aces;
  hence $53/5=10.6$ for the standard setup.
  After having solved this Question One I go on to
  certain alternative solutions and extensions,
  involving e.g.~Beta approximations. I also consider
  the distributions and means for the 2nd, the 3rd, the 4th
  occurrences of aces, with generalisations,
  where there is a Dirichlet distribution in wait for us,
  with further links to order statistics for the uniform.
  Furthermore, an apparatus is developed for obtaining
  estimators and full confidence distributions for
  applications where one knows the number $n$ of aces,
  but not the deck size $N$; and correspondingly for
  inference about the unknown population size $N$
  when $n$ is known. If you have 1000 people in a room,
  and need to interview 11 of them until you've found
  the first left-handed person, how may left-handed
  are there in the room -- here we need both an estimate
  and a clear measure of uncertainty. 

}
\end{abstract}


\section*{First the Footnote (and the Lament)}

I enjoyed working on this for two lovely summer days
on our balcony in the calendar year 2026 A.D.,
where humankind finds itself at one of its crossroads --
and where the choice \& decision have been made \& taken;
we're all part of The Manhattan Project of Our Time,
whether we lose bits of our souls and minds in the process or not.
Those times on our planet where mathematical puzzles
of this kind could be worked with,
`how many cards until I have an ace',
whether recreational or closer to higher scholarly levels,
without any laptop in the room being able to crank out
a clear answer in a minute, have passed.
No, I have {\it not} used AI for this, it hasn't yet
lured and sirened me, ``and the authors have refrained from using any AI'',
as we write in our {\it Disclosure Statement}
of Hjort og Stoltenberg (2026a, and with extended explanations
in our book 2026b; we briefly considered using the verb
`abstained'). I am willing to believe, though, that AI
rather soon will be able to give ok answers to prompts
along these lines: ``Solve the aces-in-a-deck puzzle,
and write up a six-page essay on this and its extensions
and generalisations, in the occasionally flowery literary
style of Professor N.L.~Hjort, with a bit of confidence
or Bayesian nonparametrics,  and perhaps with an indirect
pointer to two to history or literature''.
This is spellbindingly splendid -- and yet, troubling
and worrisome. It's hopeless, and we don't give in.

\section{The time until the first ace}
\label{section:intro}

The puzzle that initiated my two work-days is the following,
given as Puzzle \#64 in an impressively long series of such,
composed over many years by Professor Jostein Lillest\o l,
for the {\it Tilfeldig Gang} (word-play-ish on `Random Walk')
regular publication of the Association of Norwegian Statisticians,
was formulated as follows: 

\smallskip\noindent
{\sl `Take a deck of cards with its 52 cards, including its 4 aces.
Shuffle well, and then pick one card at a time, until
the first ace appears. What is the expected number of cards needed?
Challenge: find different solutions.'}

Let me start with the Christmas Stocking Formula, a convenient
one for what follows. It says
\beq
\begin{array}{rcl}
\displaystyle
      {3\choose 3}+{4\choose 3}+\cdots+{50\choose 3}+{51\choose 3}
   &=&\displaystyle {52\choose 4}, \\ [3mm]
\displaystyle
      {4\choose 4}+{5\choose 4}+\cdots+{51\choose 4}+{52\choose 4}
   &=&\displaystyle {53\choose 5}, 
\end{array} 
\label{eq:julestroempe} 
\eeq
etc. This combinatorial formula goes by different names.
Write down the first 7-8-9 rows in Pascal's Triangle;
sum away along a diagonal; and lo {\it\&} behold!,
the answer is found one step away in the next line.
So `hockey-stick formula' is also sufficiently appropriate,
depending on your cultural upbringing.
Put Pacal's glasses on your nose and check
e.g.~that $1+4+10+20=35$, and by summing longer you get
the first formula above. For the second you sum along the
4th diagonal downwards, with $1+5+15+35+\cdots$, etc.

\begin{small}
\begin{verbatim}
                   1
               1   2   1  
             1   3   3   1 
           1   4   6   4   1
         1   5  10  10   5   1
       1   6  15  20  15   6   1
     1   7  21  35  35  21   7   1   
Pascal's triangle, the first seven rows.
\end{verbatim} 
\end{small}

The formulae can be shown by induction, or in a direct
combinatorial fashion. One little narrative there,
verifying the formula, is as follows.
You are to place {\it five aces}, among $N+1=53$ cards,
on the table in front of you, in places 1 to 53,
which can be done in ${53\choose 5}$ ways.
If the biggest is 5: the rest can be done in ${4\choose 4}=1$
ways. If the biggest is 6: the rest can be done in ${5\choose 4}$
ways -- etc. The general version says 
\beqn
   {k\choose k}+{k+1\choose k}+\cdots+{m-1\choose k}+{m\choose k}
   ={m+1\choose k+1}. 
\eeqn

I'm returning to the deck of cards at my nearest casino
(curiously and perhaps appropriately, the son of the father
of Bayesian Nonparametrics is the world's best poker player),
and let $T$ be the number of draws until the first ace.
With $N$ cards, with $n$ aces and $m=N-n$ non-ace,
it's not hard to find the point probabilities
$f_j=\Pr(T=j)$, thinking `one card at a time, then
conditioning':
\beqn
f_1&=&{n\over N}
   ={4\over 52}, \\ 
f_2&=&{m\over N}{n\over N-1}
    ={48\over 52}{4\over 51}, \\
f_3&=&{m\over N}{m-1\over N-1}{n\over N-2}
   ={48\over 52}{47\over 51}{4\over 50}, \\
f_4&=&{m\over N}{m-1\over N-1}{m-2\over N-2}{n\over N-3}
   ={48\over 52}{47\over 51}{46\over 50}{4\over 49},    
\eeqn 
etc. With a little algebra and fiddling with the factorials,
the general formula becomes
\beq
\begin{array}{rcl} 
f_j&=&\displaystyle
  {m\over N}{m-1\over N-1}\cdots{m-(j-2)\over N-(j-2)}{n\over N-(j-1)} \\
&=&\displaystyle
  {m!\over N!} {(N-j+1)!\over (m-j+1)!} {n\over N-j+1} \\
  &=&\displaystyle
  {N-j\choose n-1}\Big/{N\choose n}, 
\end{array} 
\label{eq:generalf}
\eeq
for $j=1,2,\ldots,m+1$. As a little check we can ponder through 
\beqn
f_{47}={48\over 52}{47\over 51}\cdots{3\over 7}{4\over 6},
\quad 
f_{48}={48\over 52}{47\over 51}\cdots{3\over 7}{2\over 6}{4\over 5},
\quad 
f_{49}={48\over 52}{47\over 51}\cdots{3\over 7}{2\over 6}{1\over 5}{4\over 4}. 
\eeqn
Summing the probabilities must give us 1, which agrees with
\beqn
\sum_{j=1}^{m+1} {N-j\choose n-1}
   ={n-1\choose n-1}+{n\choose n-1}+\cdots+{N-2\choose n-1}+{N-1\choose n-1}
   ={N\choose n},
\eeqn
as with ${3\choose 3}+\cdots+{51\choose 3}={52\choose 4}$,
the first Christmas Stocking Formula in 
(\ref{eq:julestroempe}). This is actually an independent
probability based proof of that formula. 

\section{And what's the mean?}

In the puzzle we're asked about the expected value. 
Let us check with 
\beqn
\E\,(N+1-T)&=&\sum_{j=1}^{m+1} (N+1-j)f_j \\
&=&\sum_{j=1}^{m+1} (N+1-j){m!\over N!}
    {(N-j+1)!\over (m-j+1)!} {n\over N-j+1} \\
&=&n{m!\,n!\over N!}\sum_{j=1}^{m+1} {(N-j+1)!\over (m-j+1)!\,n!} \\
&=&n{1\over {N\choose n}} \Bigl[{n\choose n}
       +{n+1\choose n}+\cdots+{N\choose n}\Bigr]
   =n{N+1\over n+1},     
\eeqn 
which for the deck of cards uses 
${4\choose 4}+\cdots+{52\choose 4}={53\choose 5}$,
Christmas Stocking Formula no.~2 in (\ref{eq:julestroempe}). 
For the mean $\xi=\E\,T$ this leads to 
$N+1-\xi=n(N+1)/(n+1)$ and to the rather nice formula
\beq
\E\,T=\xi={N+1\over n+1},
\quad \hbox{which for the deck of cards means}\quad {53\over 5}=10.6.
\label{eq:themean}
\eeq 

A different solution, correlated with other insights,
uses the well-known trick that
\beqn
\E\,T=f_1+2f_2+3f_3+\cdots=S_1+S_2+\cdots, 
\eeqn 
featuring $S_j=\Pr(T\ge j)=p_j+p_{j+1}+\cdots$,
the survival probability, in the lingo of survival analysis. 
And these $S_j$ are clean enough,
\beqn
S_1=1, \quad
S_2={m\over N}={48\over 52}, \quad
S_3={m\over N} {m-1\over N-1}={48\over52} {47\over 51}, \quad
S_4={m\over N} {m-1\over N-1} {m-2\over N-2}
   ={48\over52} {47\over 51} {46\over 50}, 
\eeqn 
etc.. The general formula becomes 
\beqn
S_j={m\over N}{m-1\over N-1}\cdots{m-(j-2)\over N-(j-2)}
   ={m!\,n!\over N!} {(N+1-j)!\over (m+1-j)!\,n!}
   ={N+1-j\choose n}\Big/ {N\choose n}, 
\eeqn 
which for our favourite deck of cards means 
\beqn
S_j=\Pr(T\ge j)={53-j\choose 4}\Big/ {52\choose 4},
  \quad{\rm for\ } j=1,2,\cdots,49.  
\eeqn 
It remains to sum these -- which, again, is nice \& clean
via the Christmas Stocking:
\beqn
\E\,T={53\choose 5}\Big/{52\choose 4}={53\over 5}=10.6. 
\eeqn 
The general formula, for other decks of cards with other
subsets of aces, is as in (\ref{eq:themean}).

We note that $T$ becomes stochastically larger in $N$
and smaller in $n$, with consequences for constructing
confidence distributions below.

\begin{figure}[h]
\centering
\includegraphics[scale=0.35]{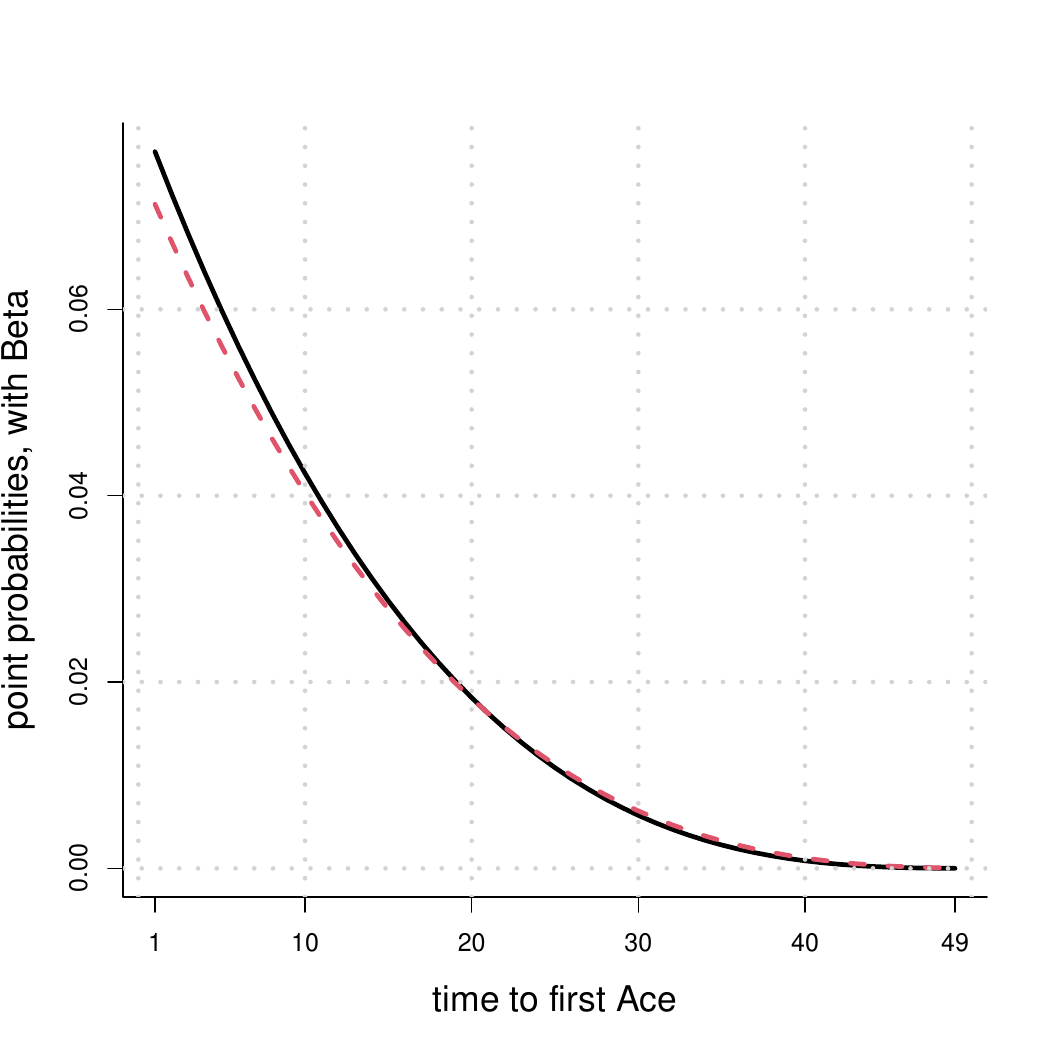}
\includegraphics[scale=0.35]{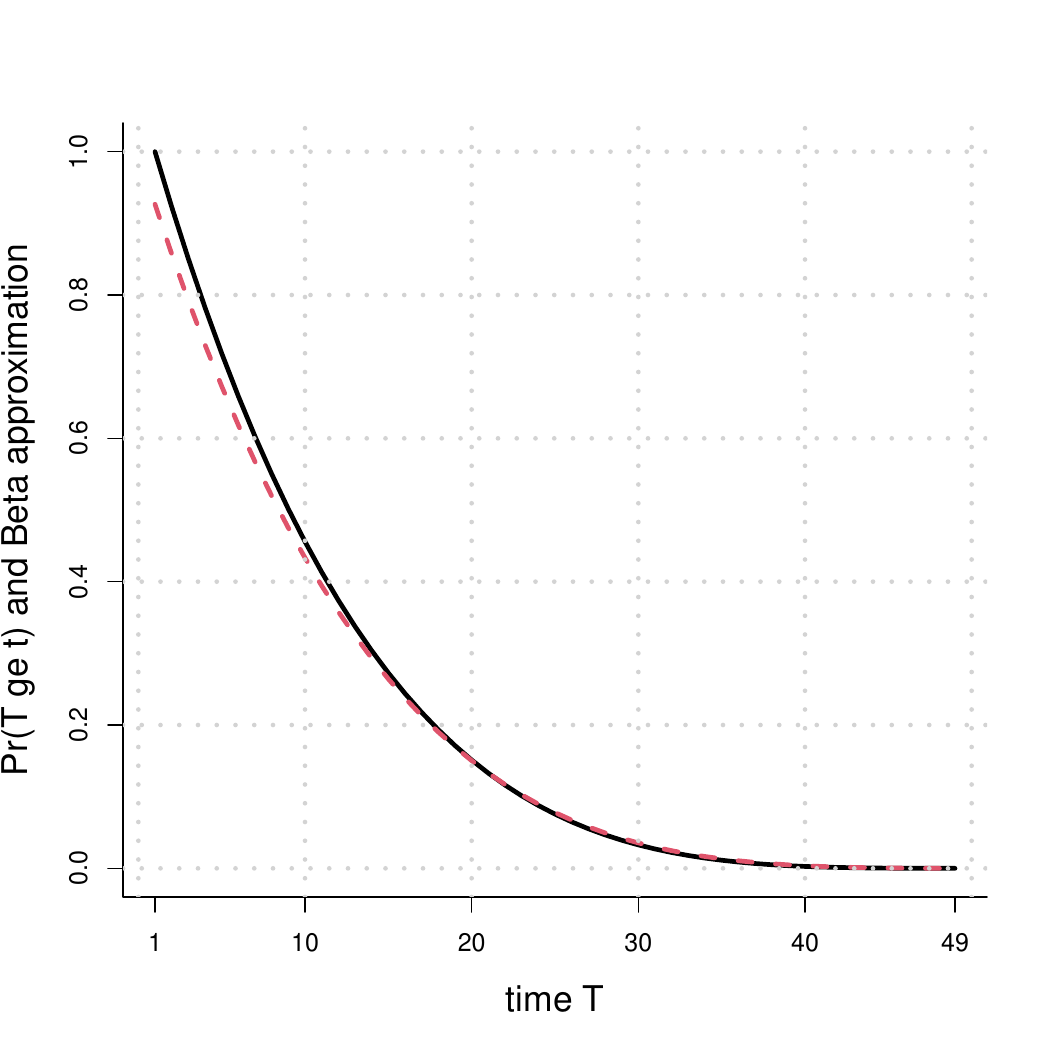}
\caption{The exact distribution of $T$, the time for
  the first ace (black, full), with the Beta approximation
  (red, dashed).
  Left: point probabilities $f_j=\Pr(T=j)$;
  right: survival probabilities $S_j=\Pr(T\ge j)$.
  The mean value for the approximation is identical
  to the exact, $53/5=10.6$.}
\label{figure:josteinA}
\end{figure}

\section{A Beta approximation}

For non-small deck of cards, as for our standard $N=52$,
the situation is `almost continuous'. Consider 
$B_N=T/(N+1)$, the normalised waiting time. Then 
\beqn
\Pr(B_N\ge x)=\Pr(T\ge (N+1)x) = {N+1-(N+1)x\choose n}\Big/{N\choose n}. 
\eeqn 
But for this we may show, via the hammer of Stirling 1730
or without, that it tends to $(1-x)^n$, for $x\in(0,1)$.
This means convergence in distribution, 
\beqn
B_N\arr_d\Bet(1,n), 
\eeqn
i.e.~a $\Bet(1,4)$, with mean value $1/5$, for our $n=4$ aces. 
Figure \ref{figure:josteinA} demonstrates that the approximation
works very well, for $(N,n)=(52,4)$.
To the left we see the exact $f_j$, along with the approximation 
\beqn
f_j^*=\Pr\{(N+1)\,\Bet(1,n)\in[j\pm\half]\}
   =\be(j/(N+1),1,n)/(N+1), 
\eeqn 
with $\be(x,1,n)=n(1-x)^{n-1}$ the Beta density,
with cumulative $\Be(x,1,n)=1-(1-x)^n$.
To the right we have the exact $S_j$, with the approximation 
\beqn
\Pr((N+1)\,\Bet(1,n)\ge j)=\{1-j/(N+1)\}^n. 
\eeqn 

Since everyone in the room knows the expected value of
the Beta, we have
\beqn
\E\,T\approx (N+1)\,\E\,\{\Bet(1,n)\}={N+1\over n+1}, 
\eeqn
once again; the mean of the approximation is the correct mean.
We may also harvest a good approximation to the variance of $T$,
from $T=(N+1)\,B_N$ and the Beta distribution. 

\section{Can we guess what $N$ or $n$ is, after having seen $T$?}

\begin{figure}[h]
\centering
\includegraphics[scale=0.35]{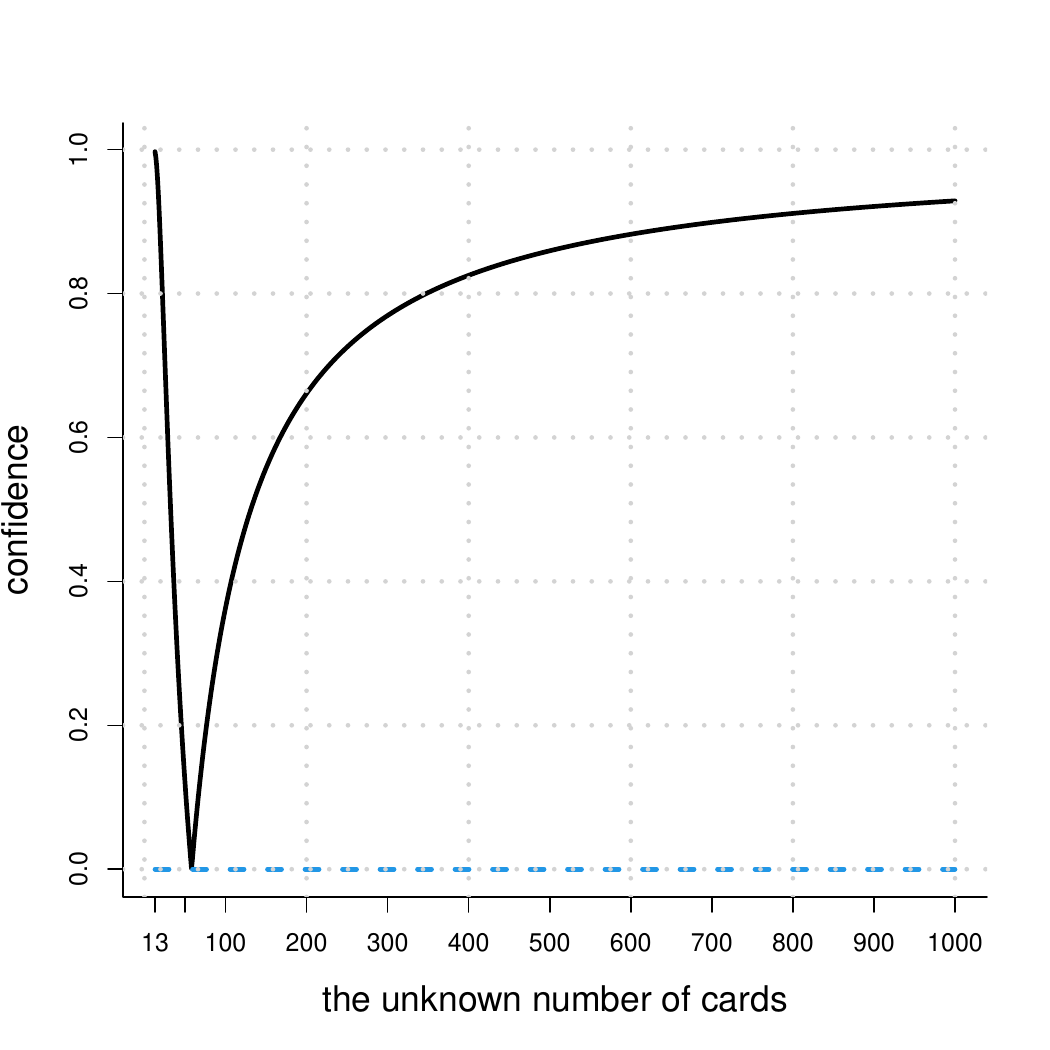}
\includegraphics[scale=0.35]{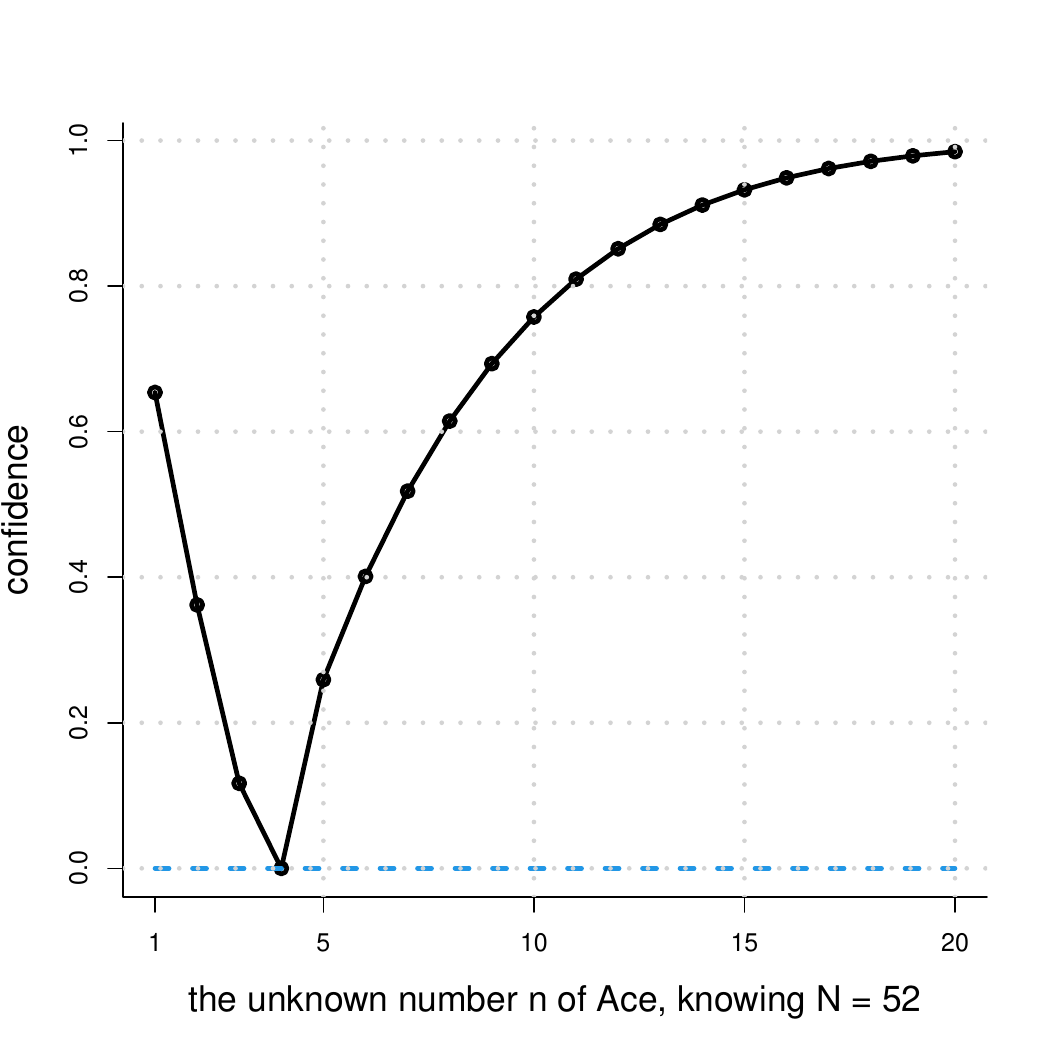}
\caption{Two confidence curves, after having needed
  $T=t_\obs=10$ draws for my first ace.
  Left: I do know there are $n=4$ aces, but estimate $N$,
  the population size. The point estimate is $\hatt N=58$,
  and the confidence quite skewed.
  Right: I do know there are $N=52$ cards, but estimate
  $n$, the number of aces there. The point estimate is
  $\hatt n=4$.}
\label{figure:josteinB}
\end{figure}

There are situations, outside casinos and card playing salons,
where one searches among individuals or objects until one
finds the first of $n$ well-defined interesting or sufficiently
important ones -- but where one does not know $N$. Similarly,
there are situations where one knows the population size $N$,
but not how many important individuals $n$ there are,
with some defining characteristics.
How can we estimate these numbers, with confidence?

\subsection*{First: given n aces, but deck size N unknown.}
Assume $T=t_\obs=10$, in such a single experiment, and that
we know there are four aces (who can forget ECh 1976, ECh 1977).
The method of moments sets $T=t_\obs$ equal to the mean value
$(N+1)/(n+1)$, yielding $\tilda N=(n+1)t_\obs -1=49$; good.

A better method, as one may show, from expected precision,
is via Schweder and Hjort (2016, Chs.~3--4) and their confidence
distributions (CDs). We do 
\beqn
C_A(N)=\Pr_N(T\ge t_\obs=10)={N+1-t_\obs\choose n}\Big/{N\choose n}
\quad {\rm for\ }N\ge N_{\rm lower}=t_\obs+n-1=13
\eeqn 
as cumulative confidence (it is monotone growing in $N$),
and from this the confidence curve 
\beqn
\cc_A(N)=|1-2\,C_A(N)|. 
\eeqn 
This is depicted in Figure \ref{figure:josteinB}, left panel.
The median confidence estimate is $\hatt N=58$,
and the curve's drastic right skewness reflects the
significant uncertainty associated with a single experiment,
All confidence intervals, for e.g.~levels 90\% or 80\%,
may be easily enough read off. 

\subsection*{Then: given N cards, with number of aces n unknown.}
The methodology pointed to above, from Schweder and Hjort (2016),
may be used here too, though with different types of curves. 
The CD here, which works `the other way',
uses that $T$ is big when $n$ is small, and takes the form
\beqn
C_B(n)=\Pr_n(T\le t_\obs=10)={N+1-t_\obs\choose n}\Big/{N\choose n}
\quad {\rm for\ }n=1,2,3,\ldots. 
\eeqn 
This is monotone growing in $n$.
Again with the $T=t_\obs=10$, as an illustration, gives the
confidence curve $\cc_B(n)=|1-2\,C_B(n)|$, shown in 
Figure \ref{figure:josteinB}, right panel, with
median estimate $\hatt n=4$. 

Methods related to those exhibited here, for aces in a deck of cards,
can be used to estimate animal abundance, and
more generally for `counting the uncounted',
as actually seen in several among the 100 Statistical Stories
in Hjort and Stoltenberg (2026b);
please read and work through 
Story \#30 (How many Clethrionomys glareoli);
Story \#31 (How many deer in the forest);
Story \#42 (How many Abel envelopes in 1902);
Story \#64 (How many were killed in Guatemala, 1978--1995). 

\section{The second time, the third time, the fourth time}

Above out attention was on $T_1$, the first time we see an ace,
and we've managed to find the point probabilities $f(t,N,n)$
from (\ref{eq:generalf}) for $t=n,n+1,\cdots,m+1$. 
But what then, for $T_2$, the second time we see an ace? 
Given $T_1=t_1$ the situation is as earlier, for the gap $T_2-T_1$,
modulo the altered parameters, which now are $N-t_1$ cards with $n-1$ aces.
Hence we have 
\beqn
\E\,\{(T_2-T_1)\midd t_1\}={N+1-t_1\over n}, 
\eeqn
which via double expectation implies 
\beqn
\E\,(T_2-T_1)={N+1\over n}-{N+1\over n+1}{1\over n}
   ={N+1\over n}\Bigl(1-{1\over n+1}\Bigr)={N+1\over n+1},
\eeqn
i.e.~the same $\xi$ as for the main formula (\ref{eq:themean}).
We do find the same for $T_3-T_2$, which for given $(T_1,T_2)=(t_1,t_2)$
must have the distribution $f(g_3,N-t_2,n-2)$. We may
state, actually without having worked hard with the underlying
probability distributions for $T_2,T_3,T_4$, that the gaps 
\beqn
G_1=T_1,\quad
G_2=T_2-T_1,\quad
G_3=T_3-T_2, \quad G_4=T_4-T_3,\ldots  
\eeqn
have the very same mean value $\xi$.
For our default deck of cards we have 
\beqn
(\E\,T_1,\E\,T_2,\E\,T_3,\E\,T_4)=(53/5,106/5,159/6,212/5)
   =(10.6,21.2,31.2,42.4),
\eeqn
which means card positions $1,2,\ldots,52,53$ partitioned
in five equally big portions. 

The gaps $G_2,G_3,\ldots$ must actually have the same marginal
distribution as the first, $G_1=T_1$. They are not independent,
but exchangeable, as de Finetti would have said (when he spoke English). 
With the notation $f(t,N,n)$ for $\Pr(T_1=t)$, with a full deck $N$
and all aces $n$ in place, as in (\ref{eq:generalf}),
one may prove, for $G_2=T_2-T_1$ marginally, that 
\beqn
f(g_2)=\Pr(T_2-T_1=g_2)=\sum_{t_1} f(t_1,N,n)f(g_2,N-t_1,n-1)=f(g_2,N,n).  
\eeqn 
We may hence see the aces times as 
\beqn
T_1=G_1,\quad
T_2=G_1+G_2,\quad
T_3=G_1+G_2+G_3,\quad
T_4=G_1+G_2+G_3+G_4,
\eeqn 
where $G_1,\ldots,G_4$ have the same distribution (but they have
negative correlations). This matches perfectly the mean
values $(\xi,2\xi,3\xi,4\xi)$ found above. 

\section{The continuous model (when the deck of cards is big)}

\begin{figure}[h]
\centering
\includegraphics[scale=0.60]{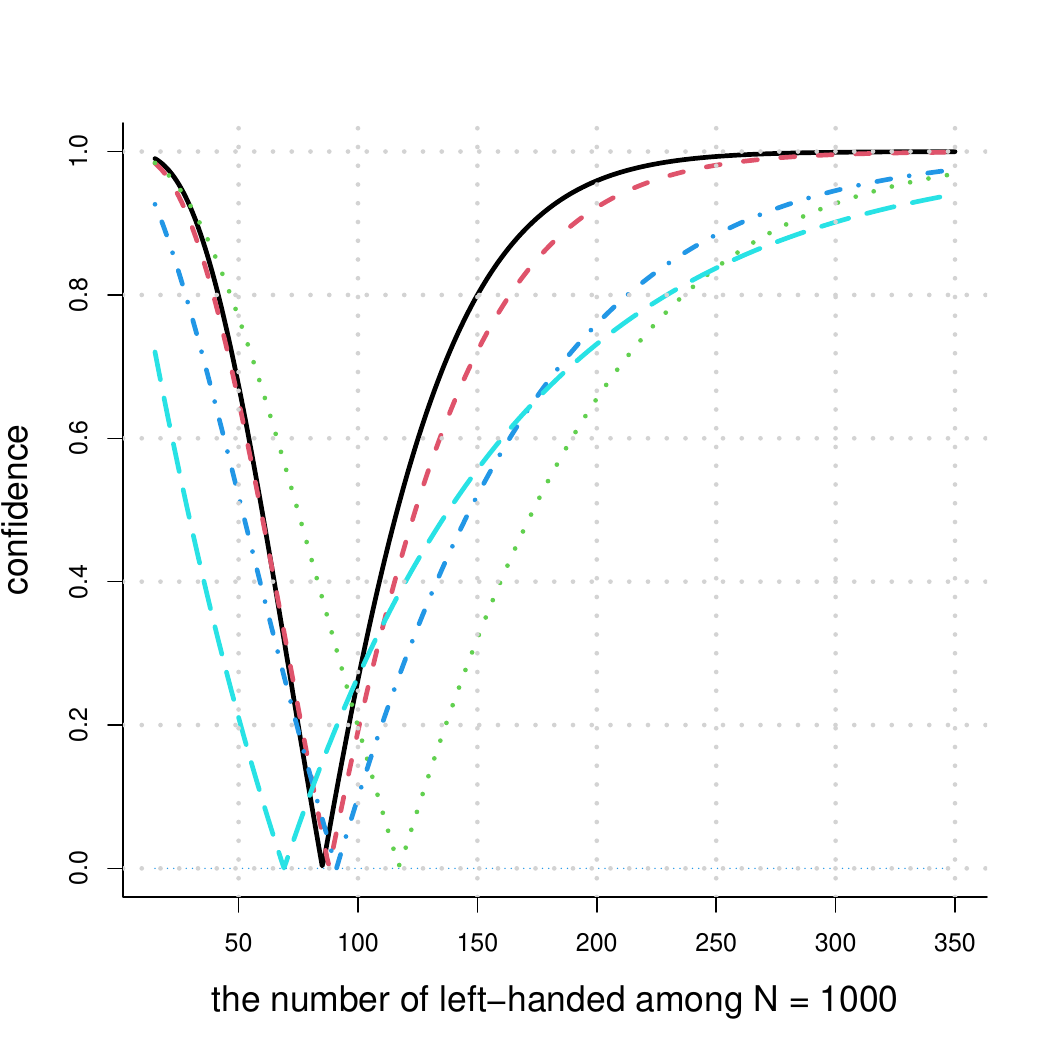}
\caption{How many left-handed are there, among $N=1000$
  persons, if I check persons one by one, and find
  left-handers as numbers $(T_1,T_2,T_3,T_4,T_5)=(10,18,22,39,50)$?
  Here are confidence curves 1, 2, 3, 4, 5, with the steadily
  growing amount of information, and where the fifth is
  the most informative one (black curve).
  The estimate is $\hatt n=85$, with 90 percent
  interval $[33,172]$.}
\label{figure:josteinC}
\end{figure}

We've managed to analyse the card situation rather accurately,
via the exact distributions for aces times
$T_1,T_2,\ldots$, etc. The answers are perhaps mildly
non-simple, but understandable and applicable, as demonstrated
above with the confidence curves $\cc_A(N)$ and $\cc_B(n)$.
Arguably, the full situation becomes nicer and easier
to analyse, mathematically and statistically, when we
stride the gap over to the continuous side.
The approximations worked out below, for the continuous
formulation, will work well, as long as $N$ is moderate
to big (surely 52 is fine) and $n$ is not big compared
to the $N$.

The key is to transform the aces times $T_1<T_2<T_3<\cdots$
to ratios 
\beqn
B_1=T_1/(N+1),\quad
B_2=(T_2-T_1)/(N-T_1),\quad
B_3=(T_3-T_2)/(N-T_2),\ldots. 
\eeqn
Each $B_j$, given the past, is a new independent variable,
of the same character, but with steadily smaller decks of cards.
Having already understood that $B_1\sim\Bet(1,n)$,
we also understand how matters turn out for the others,
given the earlier ones; when $N$ grows we must have
\beqn
B_1\sim\Bet(1,n),\,\,
B_2\midd T_1\sim\Bet(1,n-1),\,\,
B_3\midd (T_1,T_2)\sim\Bet(1,n-2),\ldots,
\eeqn
with new betas $B_j$ given the previous ones.
This may also be transformed to the scaled gaps
\beqn
D_1&=&T_1/(N+1),\\
D_2&=&(T_2-T_1)/(N+1),\\
D_3&=&(T_3-T_2)/(N+1),\ldots \\
D_n&=&(T_n-T_{n-1})/(N+1), \\
D_{n+1}&=&(N-T_n)/(N+1), 
\eeqn 
with $D_1+\cdots+D_n+D_{n+1}=1$. It's then a good exercise
in transformations of variables, complete with a Jacobi determinant,
to show from the Beta parts that 
\beqn
(D_1,D_2,\ldots,D_n,D_{n+1})\sim\Dir(1,1,\ldots,1,1), 
\eeqn 
a flat Dirichlet. But this is surely the same distribution
sa for $U_1,U_2,\ldots,U_n,1-U_n$, the order statistics
from the uniform on the unit interval. We have hence
reached the insight that
\beqn
(T_1,\ldots,T_n)\sim(N+1)(U_1,\ldots,U_n), 
\eeqn 
scaled order statistics -- and we therefore `know all details'
of the full aces-placing process.

For instance, $U_k\sim D_1+\cdots+D_k\sim\Bet(k,n+1-k)$,
with consequent
\beqn
\E\,U_k={k\over n+1},\quad
\Var\,U_k={k\over n+1}\Bigl(1-{k\over n+1}\Bigr){1\over n+2}, 
\eeqn 
which immediately gives good approximations to
means and variances for the $T_k=(N+1)U_k$.
In particular, yet again, since the Tilfeldig Gang Puzzle \#64
wished for `alternative solutions',
we do have our by now dear old $\E\,T_k=k(N+1)/(n+1)$. 

These can then be used, in ymist ways, for estimation
and inference for various `how many are there' questions.
Assume we have $N=1000$ people in a room,
and that I greet one by one, to estimate the number $n$
of left-handers. Suppose that the left-handers I meet
are nos.~$(T_1,T_2,T_3,T_4,T_5)=(10,18,22,39,50)$
(whereupon I stop interviewing the remaining 950).
What's a good estimate for the number $n$ of left-handed,
and a good confidence interval?

First, that density of the first five uniform order
statisitcs is
\beqn
h(u_1,\ldots,u_5)=1 \cdot1 \cdot1 \cdot1 \cdot1 
{n\choose 5}(1-u_5)^{n-5}
\quad {\rm for\ }u_1<\cdots<u_5<1.  
\eeqn 
From this we see that $U_5$ is sufficient. Scaling back to
the 1000 people in the room, we learn that $T_5=50$
is sufficient information. We then construct the
sufficiency based CD 
\beqn
C(n)=\Pr_n(T_5\le 50)=\Pr((N+1)\,\Bet(5,n-4)\le 50)
   =\Be(50/(N+1),5,n-4),
\eeqn 
with the Beta cumulative. 
Figure \ref{figure:josteinC} gives the confidence curve 
$\cc(n)=|1-2\,C(n)|$, with median confidence estimate 
$\hatt n=85$, along with the rather skewed 90 percent interval 
$[33,172]$. 

My students in the various courses have needed to getting
used to `extra footnotes and half-relevant digressions',
and here is one such. We may write 
\beqn
U_1=B_1,\quad
U_2=U_1+(1-B_1)B_2,\quad 
U_3=U_2+(1-B_1)(1-B_2)B_3,\ldots,
\eeqn
as a stick-breaking algorithm
(we break a stick in two pieces;
then the remaining part in two; etc.),
with independent
$B_1\sim\Bet(1,n)$, $B_2\sim\Bet(1,n-1)$, $B_3\sim\Bet(1,n-2)$, etc. So 
\beqn
U_k=D_1+\cdots+D_k, \quad {\rm with\ }
D_i=(1-B_1)\cdots(1-B_{i-1})B_i,  
\eeqn
which implies a representation for the $U_k$,
and in their turn again our friends $T_k$ from
the deck of cards, in terms of
products of independent Beta variables.
This is a finite-sum-product representation
for a full Dirichlet vector,
related later on to the Sethuraman infinite-sum-product
representation of the Dirichlet process,
a key friend of scholars in Bayesian Nonparametrics.

\section*{References}

\def\ref#1{{\noindent\hangafter=1\hangindent=20pt
  #1\smallskip}}          
\parindent0pt
\baselineskip11pt
\parskip3pt 

\begin{description}

\item N.L.~Hjort, C.C.~Holmes, P.~M\"uller, S.G.~Walker (2010).
  {\it Bayesian Nonparametrics.}
  Cambridge University Press. 

\item N.L.~Hjort (2008).
  The correlation between mean and median:
  essay concerning Puzzle \#23 in Tilfeldig Gang no.~1, 
  {\it Tilfeldig Gang} no.~2, 22--25. 

\item N.L.~Hjort (2026). Counting the uncounted:
  How many were killed in Guatemala, 1978-1995?
  {\sl Proceedings of the International Workshop
    on Statistical Modelling}, Oslo, July 2026, invited talk. 

\item N.L.~Hjort, E.Aa.~Stoltenberg (2026a).
  Probability Proofs for Stirling (and More):
  The Ubiquitous Role of $\sqrt{2\pi}$.  
  {\it American Statistician}, {\bf 80}, 405--412.
  
\item N.L.~Hjort, E.Aa.~Stoltenberg (2026b).
  {\it Statistical Inference: 600 Exercises and 100 Stories.}
   Cambridge University Press. 
 
\item T.~Schweder, N.L.~Hjort (2016).
  {\it Confidence, Likelihood, Probability.
    Statistical Inference With Confidence Distributions.}
  Cambridge University Press. 
  
\end{description}

\end{document}